\documentclass{desyproc}

\def\lesssim{\ \hbox{\raise 2pt \hbox{$<$} \kern -13pt
                     \lower 3pt \hbox{$\sim$}}\ }
\def\greatersim{\ \hbox{\raise 2pt \hbox{$>$} \kern -13pt
                     \lower 3pt \hbox{$\sim$}}\ }
\def\cascade{{\sc Cascade}}
\def\herwig{{\sc Herwig}}
\def\pythia{{\sc Pythia}}

\begin{document}
\title{Parton showering effects in central  heavy-boson\\ 
\hspace*{4.7 cm}  hadroproduction}

\author{\hspace*{2 cm}  {\slshape M. Deak$^1$, A. Grebenyuk$^2$, 
F. Hautmann$^3$,  H. Jung$^{2,4}$  and  K. Kutak$^2$ }\\[1ex]
\hspace*{5.3 cm} $^1$IFT, University of Madrid\\
\hspace*{3.8 cm} $^2$Deutsches Elektronen Synchrotron, Hamburg\\
\hspace*{3.3 cm} $^3$Theoretical Physics Department, University of Oxford\\
\hspace*{4.0 cm} $^4$Physics Department,  University of Antwerp}


\desyproc{DESY-PROC-2009-xx}
\acronym{EDS'09} 
\doi  

\maketitle

\begin{abstract}
If  large-angle multigluon radiation  contributes   significantly   
to parton  showers associated with heavy boson production at the LHC, 
appropriate  parton branching methods are required for 
realistic  Monte Carlo simulations of final states.  We report 
on a  study   illustrating  such effects  in the case of central 
scalar-boson  production. We comment on the possible impact of 
such studies on the 
modeling of multi-parton interactions. 
\end{abstract}

\section{Introduction}

Final states  containing  heavy bosons  and    jets   will be crucial  
in a number of     experimental searches at the Large Hadron Collider. 
Phenomenological  analyses   will   rely both on perturbative 
fixed-order  calculations and on    parton shower   Monte Carlo generators 
for   a  realistic  description of the structure  of   these  events.    

Due  to the presence of multiple hard scales and the large phase space 
opening up at LHC  energies,   the  treatment of these  final  states   is  
 potentially  sensitive to complex dynamical effects in the QCD showers 
 accompanying the   events.   In the case of vector bosons 
  it has been pointed out~\cite{pz_perugia} that the 
 treatment of parton showers,  and 
in particular  of the recoils in the  shower, 
 is essential for a proper description 
of the  W/Z   $p_T$ spectrum. 
This affects the amount of multi-parton 
   interactions~\cite{pz_perugia,sjo09,giese,gustafstalk} 
needed to 
describe the events. On the other hand, 
parton showers which are not ordered in transverse momentum could 
also considerably contribute to 
what is typically associated with the underlying event.  
In the case of vector bosons  this may be     
    relevant for  early phenomenology at the LHC, as  the 
  possible  broadening of W and Z $p_{T}$ 
 distributions~\cite{cpyuan1}    affects the 
  use of  these processes      as luminosity monitor~\cite{mandy}. 

For scalar  boson production, the role of corrections to  
 transverse-momentum  ordered showers on the structure of final states was 
 considered in~\cite{lonn,jung04}  in terms of the 
 heavy-top  effective theory  matrix elements associated with the unintegrated 
 gluon density~\cite{higgs02}.   In this article we report on 
 ongoing studies~\cite{nastjaetal} of   mini-jet radiation  accompanying scalar 
 boson production  in the central region at the LHC. It is appropriate to 
  consider  this issue    in view of the progress  in  the 
  quantitative  understanding of  
  unintegrated gluon  contributions in multi-jet final  states~\cite{hj_ang}.   
 In the case of scalar bosons as well,     such  studies   have 
 implications on  the   role of  multi-parton interactions~\cite{sjo09,giese}  
   in the    evolution of the  initial  state shower.

We start in Sec.~2  with a brief discussion summarizing aspects of  
corrections to collinear-ordered showers  and the role of    recent  jet-jet 
correlation   measurements.   In Sec.~3 we  consider   the application of  
parton showers not ordered in transverse momentum to the case of 
central scalar-boson 
hadroproduction.  We conclude   in Sec.~4.

\section{Corrections to 
 collinear  showers  and  jet   correlations}

In this section we  briefly  discuss    effects  of high-energy corrections to 
collinear parton showers  on  hadronic final states  with multiple jets. 

Let us recall that    
the    branching algorithms underlying the  most commonly  used   
 shower Monte Carlo  event generators~\cite{mc_lectures,qcdbooks}  
are  based on 
collinear evolution of jets developing, both ``forwards"  and 
``backwards",   from the hard 
event~\cite{CSS},  supplemented  (in the   case of certain generators)   by 
suitable   constraints for 
angularly-ordered phase space~\cite{bryan86}.  
The angular  constraints  are 
 designed   to 
take  account of   coherence effects  from  
multiple soft-gluon emission~\cite{bryan86,bcm,dok88}.

The    main new effect  one  observes 
when trying   to  push   this picture to higher and higher energies   
 is that  soft-gluon insertion  
rules~\cite{bcm,dok88}  based on  
eikonal emission currents~\cite{griblow,jctaylor}  
are modified in the high-energy, multi-scale region by terms that 
depend on the total transverse momentum transmitted down the 
initial-state parton decay chain~\cite{skewang,hef90,mw92}. 
As a result,  the physically relevant 
distribution to describe initial-state  showers  
becomes  the analogue  not so much 
of an  ordinary parton density but rather of an   ``unintegrated" parton density, 
dependent on both longitudinal and transverse 
momenta.\footnote{See~\cite{acta09} for  recent  reviews  of unintegrated 
pdfs.   Aspects of u-pdfs 
  from the standpoint  of  QCD high-energy factorization are 
discussed in~\cite{hef}.  Associated 
 phenomenological aspects  are discussed 
  in~\cite{jung04,acta09,bo_andothers},   and references 
 therein;  see~\cite{heralhc,jadach09,wattetal} for recent new work.  
  The  papers in~\cite{jcczu}  
 contain   first   discussions  of   a   more  general,   
nonlocal operator formulation    of u-pdfs applied to 
 parton showers    beyond leading order.} 

The next  observation concerns the structure of virtual 
corrections.  Besides   Sudakov form-factor  effects  included in    standard  
shower algorithms~\cite{mc_lectures,qcdbooks},  one needs  in general 
virtual-graph terms to be incorporated in 
transverse-momentum dependent (but universal)   splitting 
functions~\cite{skewang,jcczu,ch94,fhfeb07}   
in order to take  account of    gluon coherence  not only for 
collinear-ordered emissions but also in the non-ordered region that 
opens up at high $\sqrt{s} / p_\perp$. 

   These  
 finite-k$_\perp$  corrections  to parton branching 
 have important implications   for  multiplicity distributions  and 
   the structure of  angular correlations in   final states  with high 
   multiplicity.    
Refs.~\cite{hj_ang,hjradcor}  analyze     
examples  of  such effects in  the case of 
di-jet  and 3-jet production in  
$ep$~\cite{zeus1931,aktas_h104} and  
$ p {\bar p}$~\cite{d02005}  collisions.  
In particular,  the  accurate  measurements~\cite{zeus1931}  of 
azimuthal  and transverse-momentum correlations  
are compared  with   results  from collinear shower 
 (\herwig~\cite{herwref}) and k$_\perp$-shower  (\cascade~\cite{jung02}). The 
region of  large azimuthal separations between the leading jets,  
$\Delta \phi \sim  180 ^o$,  is   dominated by    soft gluon 
emission effects,  while   the region of 
small azimuthal separations,  down to $\Delta \phi \sim 30 ^o$,  is driven by hard 
parton radiation, thus offering a  significant 
 test of the quality of hard to semi-hard parton 
 showers over the full region of phase space. The 
 description of the    angular correlation measurements    by the   
 k$_\perp$-shower   
   is good,  and  provides  
 confidence on the  wider applicability of the method 
  for  multi-jet  processes. Results based on collinear 
 parton showers   (\herwig) 
 cannot describe the shape of the $\Delta \phi$ distribution.

\begin{figure}[htb]
\vspace{50mm}
\includegraphics{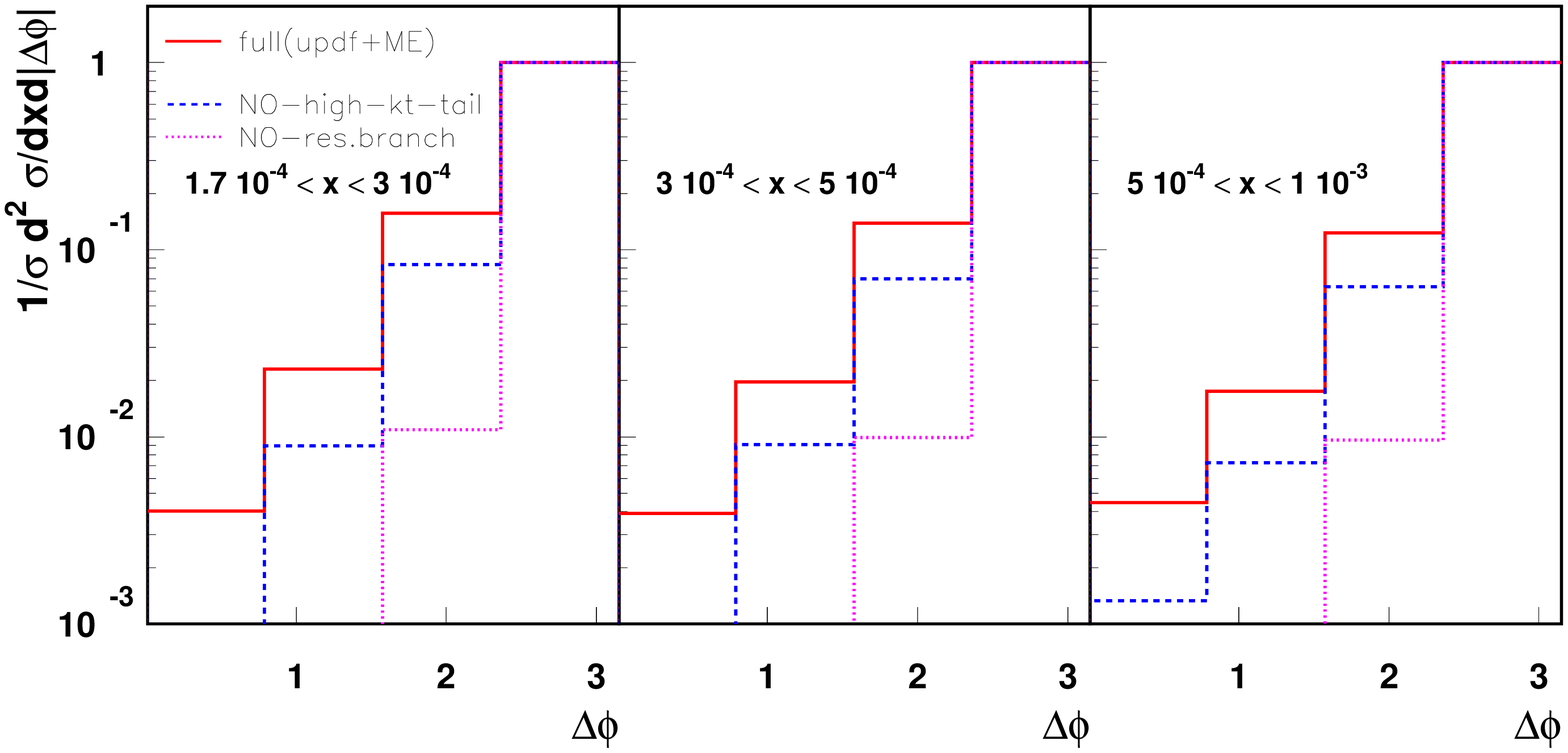}
\caption{The dijet azimuthal distribution~\protect\cite{hjradcor} 
    normalized to the
back-to-back cross section:
(solid red)  full result
(u-pdf $\oplus$ ME); (dashed blue) no finite-k$_\perp$ 
correction in ME
 (u-pdf $\oplus$ ME$_{collin.}$);
(dotted violet) u-pdf with no resolved branching.
}
\label{fig:ktord}
\end{figure}

The      k$_\perp$-shower    predictions   
involve   both  
transverse-momentum dependent   pdfs  and  
  matrix elements.
Fig.~\ref{fig:ktord}~\cite{hjradcor} 
 illustrates  the relative contribution of these   
different components  to  $ep$ di-jet cross sections 
showing different approximations to the
azimuthal dijet distribution normalized to the
back-to-back cross section. The solid
red curve is   the full  result~\cite{hj_ang}. 
  The dashed blue curve is  obtained
from the same unintegrated pdf's but
by taking the collinear approximation in
the hard matrix element.    The dashed curve  
drops much faster than the full result as $\Delta \phi$ decreases, 
indicating  that  the
high-k$_\perp$   component  in the  ME~\cite{ch94}   
is necessary  to describe 
jet correlations      
for small $\Delta \phi$.   
The dotted (violet) curve is  the result 
  obtained from the    
unintegrated pdf 
without any resolved branching.  
 This   represents   the contribution of 
the intrinsic     distribution only, 
corresponding   
 to nonperturbative, predominantly   
 low-k$_\perp$   modes.  That is,  in the dotted (violet) curve one retains 
an  intrinsic      k$_\perp$   $\neq 0$ but no  effects of coherence. We see  
 that the resulting jet correlations in this case are down by an order of magnitude. 
The inclusion of  the  perturbatively computed  high-k$_\perp$ 
 correction         distinguishes the  calculation~\cite{hj_ang}  
 from other  shower approaches   
 that include transverse momentum 
dependence in the  pdfs but not  in the  matrix elements,  see e.g.~\cite{hoeche}.

The corrections to collinear showers 
 described above  embody      the 
 physics of the unintegrated gluon 
density and associated hard matrix elements. 
Besides jet-jet correlations,   these corrections  will 
 affect the structure of     
final states  associated with  heavy mass production.  In the next section we 
consider    implications of the unintegrated gluon density and 
noncollinear contributions to showering on the 
jet  activity  accompanying production of heavy scalars in the 
central region at the LHC.

\section{Central scalar boson production at the LHC}

\begin{figure}[htb]
\vspace{49mm}
\includegraphics{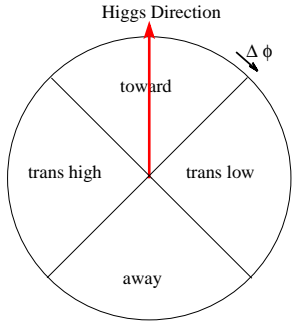}
\caption{Different regions in $\phi$ with respect to the Higgs direction.} 
\label{fig:uehiggs}
\end{figure}

To study the effect of  noncollinear  parton showers and 
its contribution to the underlying event, Ref.~\cite{nastjaetal}  
 investigates a gluon induced process 
which produces a color singlet scalar system in the final state, here  $gg \to h^0$. 
  We consider   radiation associated with  standard model Higgs boson  production,   
 following the CDF analysis of 
the underlying event~\cite{Affolder:2001xt}. 
  As shown in Fig.~\ref{fig:uehiggs}, 
the direction of the Higgs boson in the azimuthal plane 
defines the origin of the system,  
    and four 
regions in the azimuthal plane are 
defined.  

\begin{figure}[htb]
{\includegraphics[width=0.49\columnwidth]{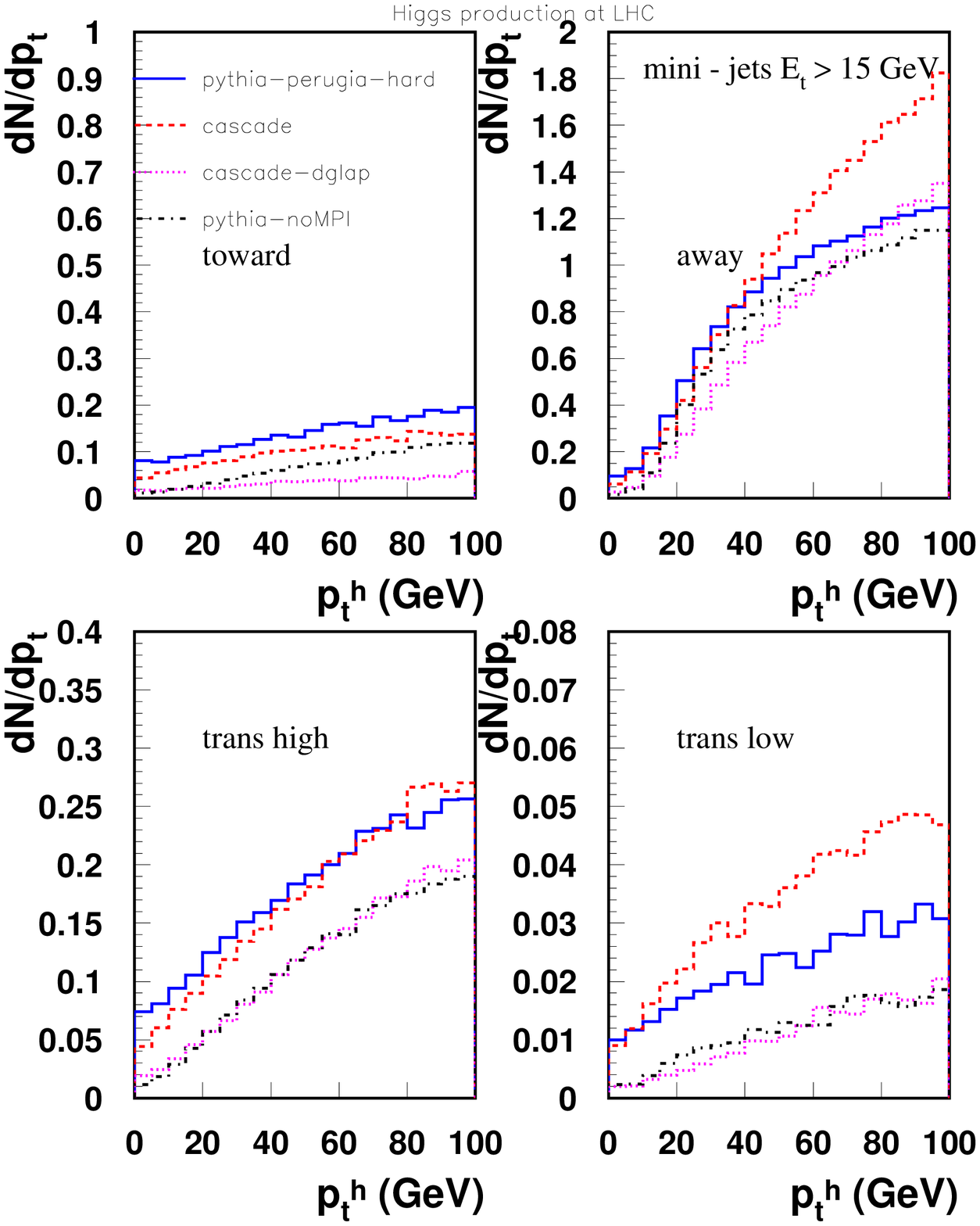}}
{\includegraphics[width=0.49\columnwidth]{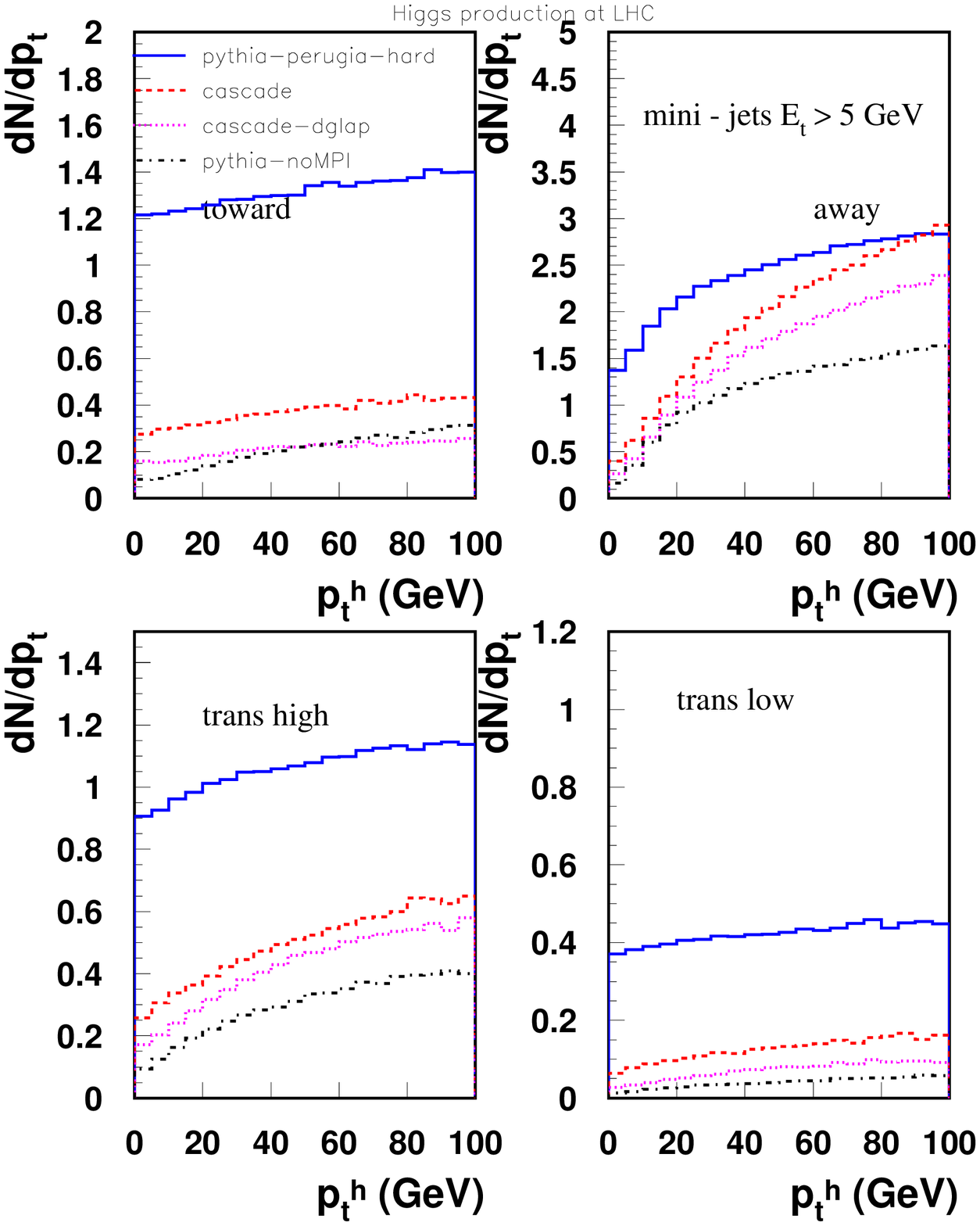}}
\caption{Multiplicity of jets  as a function of the transverse momentum of the Higgs in 
different regions of $\phi$ as predicted  from 
k$_\perp$-shower 
(\protect\cascade) and   collinear shower (\pythia). 
Shown is also the prediction using \cascade\ in collinear mode (cascade-dglap).
The left figure shows results 
 for mini-jets  with $E_T>15$ GeV, the right figure for 
 $E_T>5$ GeV.
}\label{Fig:UEHiggsCascade}
\vspace*{-0.2cm}
\end{figure}

 Fig.~\ref{Fig:UEHiggsCascade}~\cite{nastjaetal}  
 shows results for  the average multiplicity for mini-jets 
with $E_t > 15$ GeV and with $E_t > 5$ GeV 
at LHC energies ($\sqrt{s} = 14$ TeV) in the 4 
different regions of $\phi$ as a function of the Higgs transverse momentum. 

The predictions  of  the k$_\perp$-shower Monte Carlo  
generator \cascade~\cite{jung02}  are  compared with 
predictions from \pythia~\cite{pz_perugia}.  For 
comparison, \cascade\ is also run in collinear 
mode (cascade-dglap), with the off-shell matrix 
element~\cite{higgs02} replaced by the 
on-shell approximation and the parton showers are 
evolved with the one-loop splitting function and an upper 
restriction on the transverse momentum $p_t < \sqrt{m^2_h + p_{t\;h}^2}$.
For mini-jets with $E_T>15$ GeV 
\cascade\ in collinear  mode reproduces     the 
 prediction of \pythia\ without multiparton interactions in both transverse regions. 
The full \cascade\  run    
gives  higher activities in the transverse 
as well as in the toward regions, and is close or larger than   
the prediction of \pythia\ including multiparton interactions.  In the 
away region the slope is steeper than predicted from \pythia. Lowering the 
transverse momentum cut of the mini-jets  to $E_T>5$ GeV, \cascade\ still 
predicts a 
larger
 multiplicity than \pythia\ without multiparton interactions, but 
  falls clearly below the prediction including multiparton interactions. 
This illustrates the onset of hard perturbative contributions from the 
parton showers, which are simulated in \pythia\   
with multiparton interactions.

The phase space region where soft radiation plays a significant role is the region of 
minimal transverse energy in the $\phi$ plane. This is the region where   
multiparton interactions should be visible. We have also 
studied the multiplicity of charged particles (with $p_t > 150$ MeV) in the process 
$gg \to h^0$.  The study  in~\cite{nastjaetal}  indicates  that 
the result from \pythia\  
 including multiparton interactions is  above the result  from 
 \cascade, 
 however the multiplicity predicted from \cascade\   
 is significantly larger than the one predicted by \pythia\   
 without multiparton interactions.  The  result  using full 
 unintegrated-pdf  evolution 
 shows significantly more activity in all regions. Thus even for the soft contribution 
 in the charged particle multiplicity the treatment of parton showers is important.

In the region of minimum bias events, elastic and soft diffractive  processes 
will also play a role.  This is not (yet) implemented in   \cascade. 

More details of this study will be reported  in a forthcoming publication. 
We note  that the effects described above  can  
influence  the description of  soft  
underlying events and  minijets~\cite{pz_perugia,gustafstalk}   
as well as  the  use of exclusive scalar production channels~\cite{lonn}.

\section{Outlook}

The production of 
final states  containing  heavy bosons  and  multiple   jets   will be characterized at the 
 LHC by   the large phase space 
 opening up at high center-of-mass energies, and  
   the presence of   multiple  hard scales, possibly widely disparate from each other. 
This brings in  potentially large perturbative corrections  
to    hard-scattering  events   
and potentially new effects 
in the  parton-shower  components of  the  process.   
 
 If   large-angle multigluon radiation  gives   significant  contributions 
to the QCD   showers   accompanying heavy  boson  
production at  the LHC,     
appropriate    generalizations  of     parton branching  methods  
are required.  In this study we have considered  jet radiation 
associated with  heavy scalars produced centrally at the LHC, 
and we have described applications of 
 transverse-momentum dependent  
 kernels~\cite{lonn,jung04,higgs02,hef,bo_andothers}  for parton showering. 
We have focused on  associated minijet distributions  
and discussed  a comparison of showering effects with multi-parton 
interactions effects. 

This study lends itself to extensions in several directions. 
First,  we have considered here mini-jet and  effects that could  be associated  with  
the underlying event. However, the approach is  much 
 more general (see e.g. discussions in~\cite{hj_ang,acta09,heralhc,jcczu})  
    and could be used to  investigate  hard radiation as well. 
 
 Next,  we have considered scalar  boson  production
which is dominated by the physics of  initial-state  gluonic  showers, 
expressible in terms of   unintegrated  gluon densities.    
But   treatments of 
quark contributions to showers at unintegrated level   
are   also   being worked on (see e.g.~\cite{heralhc,jadach09,wattetal}).       
In this respect,   
theoretical results for splitting kernels~\cite{ch94} already  applied to  inclusive 
phenomenology  can  also be of use in  calculations  for  exclusive  
final states~\cite{hannes_curr}.
This will have    
direct  applications  to   parton showers in  vector boson production.

Further, relevant  areas  of experimental studies    will involve 
    jet physics  in the forward rapidity region~\cite{denter09} at the LHC.  
  In this article we have  limited ourselves to considering 
  production processes 
in the  central rapidity region.  Note  that   
techniques are being developed~\cite{deak09}      to  allow one to also 
 address    multi-particle   hard processes   at   forward rapidities.  

\vskip 1 cm 

\noindent {\bf   \Large  Acknowledgments} 

\vskip 0.4 cm 

\noindent We thank the organizers for the kind invitation and for the excellent organization 
of the meeting.


\begin{footnotesize}




\begin{thebibliography}{99}



\bibitem{pz_perugia}
      P.Z.~Skands, arXiv:0905.3418   [hep-ph]  in Proc.\  Perugia Workshop (2008).   

\bibitem{sjo09}
      R.~Corke and   T.~Sj{\" o}strand, arXiv:0911.1909 [hep-ph].  
      
       
\bibitem{giese}
        M.~B{\" a}hr, S.~Gieseke and M.~Seymour,       JHEP {\bf 0807} (2008) 076.            

\bibitem{gustafstalk} 
         G.~Gustafson,  talk at  Desy  Workshop, 
         Hamburg, March 2007;     G.~Gustafson, L.~L{\" o}nnblad and G.~Miu, 
         JHEP {\bf 0209} (2002) 005. 
    
    
\bibitem{cpyuan1} 
      F.I.~Olness, talk at  HERA-LHC Workshop,  CERN,  May 2008;  
      S.~Berge, P.M.~Nadolsky,  F.I.~Olness and C.P.~Yuan, hep-ph/0508215. 



\bibitem{mandy} 
      A.M.~Cooper-Sarkar,  arXiv:0707.1593 [hep-ph].




\bibitem{lonn}
    L.~L{\" o}nnblad and M.~Sj{\" o}dahl,  
    JHEP {\bf 0402} (2004) 042. 



\bibitem{jung04}
       H.~Jung,   Mod.\  Phys.\  Lett.\  A {\bf 19}   (2004) 1. 



\bibitem{higgs02}
        F.~Hautmann,  Phys.\ Lett.\ B {\bf 535} (2002) 159.  



\bibitem{nastjaetal}
        M.~Deak, A.~Grebenyuk, F.~Hautmann, H.~Jung and K.~Kutak, in  
        Proc.\  DIS09 Workshop (Madrid, 2009).     





\bibitem{hj_ang} 
       F.~Hautmann and H.~Jung,     
       JHEP {\bf 0810} (2008) 113. 




\bibitem{mc_lectures}
     B.R.~Webber, CERN Academic Training Lectures (2008). 

\bibitem{qcdbooks}
             R.K.~Ellis, W.J.~Stirling and B.R.~Webber,   
             {\em  QCD and collider physics}, CUP 1996;  
                        Yu.L.~Dokshitzer, V.A.~Khoze, A.H.~Mueller and S.I.~Troian, 
             {\em Perturbative QCD},  
              Ed.~Frontieres, Gif-sur-Yvette (1991).

\bibitem{CSS}
      J.C.~Collins, D.E.~Soper and G.~Sterman, Adv.\  Ser.\  Direct.\  High  Energy 
      Phys.\  5 (1988) 1. 
       
\bibitem{bryan86} 
           B.R.~Webber,    Ann.\ Rev.\ Nucl.\ Part.\  Sci.\ {\bf 36} (1986) 253.  

\bibitem{bcm}
      A.~Bassetto, M.~Ciafaloni and G.~Marchesini, 
      Phys.\  Rept.\  {\bf 100}  (1983) 201. 


\bibitem{dok88}
               Yu.L.~Dokshitzer, V.A.~Khoze, A.H.~Mueller and S.I.~Troian,  
               Rev.\   Mod.\   Phys.\   {\bf 60}  (1988)  373.  

\bibitem{griblow}
                V.N~Gribov, Sov.\ J.\  Nucl.\ Phys.\  {\bf 5} (1967)  399; 
                F.E.~Low, Phys.\ Rev.\ 
                 {\bf 110} (1958) 974. 



\bibitem{jctaylor}
             J.~Frenkel and J.C.~Taylor, 
                 Nucl.\ Phys.\ {\bf B246} (1984) 231; 
             R.~Doria, J.~Frenkel and J.C.~Taylor, 
                 Nucl.\ Phys.\ {\bf B168} (1980) 93.  


\bibitem{skewang}
     M.~Ciafaloni, Nucl.\ Phys.\ {\bf B296} (1988)  49.


\bibitem{hef90} 
     S.~Catani, M.~Ciafaloni and F.~Hautmann,   
     Phys.  Lett.  B{\bf 242}  (1990) 97. 
     

\bibitem{mw92}
        G.~Marchesini and B.R.~Webber,
      Nucl.\ Phys.\ {\bf B386} (1992) 215.  




\bibitem{acta09}
          F.~Hautmann,             Acta  Phys.\  Polon.\  B  {\bf 40} (2009) 2139; 
  F.~Hautmann and H.~Jung,    Nucl.\  Phys.\  Proc.\  Suppl.\  {\bf 184} (2008) 64  
         [arXiv:0712.0568 [hep-ph]]. 

\bibitem{hef} 
     S.~Catani, M.~Ciafaloni and F.~Hautmann,        
         Nucl. Phys.  B{\bf 366} (1991) 135;   
       Phys.\ Lett.\  {\bf B307}  (1993) 147.

\bibitem{bo_andothers} 
          J.R.~Andersen  {\it et al.},   Eur.\ Phys.\ J.\  C {\bf 48} (2006) 53; 
          B.~Andersson {\it et al.},       Eur.\ Phys.\ J.\  C {\bf 25} (2002)   77.         


\bibitem{heralhc}
     H.~Jung {\it et al.},    Proceedings of 
     the Workshop ``HERA and the LHC", arXiv:0903.3861 [hep-ph].


\bibitem{jadach09}
        S.~Jadach   and  M.~Skrzypek,  arXiv:0905.1399 [hep-ph];  arXiv:0909.5588 [hep-ph]. 

\bibitem{wattetal}
       A.D.~Martin,        M.G.~Ryskin and  G.~Watt,    arXiv:0909.5529 [hep-ph].


\bibitem{jcczu}
     J.C.~Collins and X.~Zu, JHEP {\bf 0503} (2005) 059; 
     J.C.~Collins,     Phys.\ Rev.\ D {\bf 65}    (2002)   094016; 
     J.C.~Collins and F.~Hautmann,       JHEP {\bf 0103} (2001) 016.  



\bibitem{ch94}     
     S.~Catani and F.~Hautmann,   
     Nucl. Phys. B{\bf 427} (1994) 475;       Phys.  Lett. B{\bf 315}  (1993) 157. 




\bibitem{fhfeb07}
     F.~Hautmann,   Phys.\ Lett.\  B {\bf  655} (2007) 26.



\bibitem{zeus1931}
    S.~Chekanov {\it et al.}  [ZEUS Collaboration],  
    Nucl.\ Phys.\  B {\bf 786} (2007) 152
    [arXiv:0705.1931 [hep-ex]].


\bibitem{aktas_h104}
    A.~Aktas {\it et al.}  [H1 Collaboration],
    Eur.\ Phys.\ J.\  C {\bf 33} (2004) 477 
    [arXiv:hep-ex/0310019].   



\bibitem{d02005}
       V.M.~Abazov  {\it et al.}  [D0 Collaboration], 
       Phys.\ Rev.\ Lett.\ {\bf 94}  (2005) 221801 
       [arXiv:hep-ex/0409040]. 

\bibitem{hjradcor}
     F.~Hautmann and H.~Jung, 
     arXiv:0804.1746 [hep-ph], 
     in Proceedings of the 
     8th International Symposium on Radiative 
     Corrections {\footnotesize RADCOR2007}; 
         arXiv:0808.0873 [hep-ph]. 

\bibitem{herwref}
  G.~Corcella {\it et al.},
  JHEP {\bf 0101} (2001) 010.

\bibitem{jung02}
       H.~Jung,   
       Comput.\ Phys.\ Commun.\  {\bf 143} (2002) 100. 

 
       
\bibitem{hoeche}
      S.~H{\" o}che, F.~Krauss and T.~Teubner, 
      Eur.\  Phys.\   J.\  C{\bf 58}  (2008)   17. 

 


\bibitem{Affolder:2001xt}
          CDF   Coll.,             Phys.\ Rev.\ D {\bf 65}    (2002) 092002.


\bibitem{hannes_curr}
          H.~Jung     {\it et al.},   in progress.   

\bibitem{denter09} 
   D.~d'Enterria, arXiv:0911.1273 [hep-ex]. 
       

\bibitem{deak09}
      M.~Deak    {\it et al.},    arXiv:0908.0538 [hep-ph];   arXiv:0908.1870 [hep-ph]. 



\end{thebibliography}
%

\end{footnotesize}


\end{document}